\documentclass{anstrans}
\title{Reducing Spatial Discretization Error with Linear Discontinuous Source Tilting in\\Iterative Quasi-Monte Carlo for Neutron Transport}
\author{Samuel Pasmann$^1$, Ilham Variansyah$^2$, C.T. Kelley$^3$, and Ryan G. McClarren$^1$}

\institute{
 $^1$University of Notre Dame, Department of Aerospace and Mechanical Engineering, Fitzpatrick Hall, Notre Dame, IN 46556, spasmann@nd.edu
 \and
 $^2$Oregon State University, Department of Nuclear Science and Engineering, 1791 SW Campus Way, Corvallis, OR
 \and
  $^3$North Carolina State University, Department of Mathematics, 3234 SAS Hall, Box 8205, Raleigh NC 27695
}

\usepackage{graphicx} % allows inclusion of graphics
\usepackage{booktabs} % nice rules (thick lines) for tables
\usepackage{microtype} % improves typography for PDF

\begin{document}
%%%%%%%%%%%%%%%%%%%%%%%%%%%%%%%%%%%%%%%%%%%%%%%%%%%%%%%%%%%%%%%%%%%%%%%%%%%%%%%%
\section{Introduction}
\subsection{The Iterative Quasi-Monte Carlo Method}
Recently, iterative Quasi-Monte Carlo (iQMC) was introduced as a new method of neutron transport which combines deterministic iterative methods and quasi-Monte Carlo simulation for more efficient solutions to the neutron transport equation \cite{pasmann2022quasi}. Primary advantages of iQMC include a vectorized multigroup scheme, $O(N^{-1})$ solution convergence, and the use of advanced iterative Krylov solvers, including GMRES and BiCGSTAB, which converge with far fewer iterations than the standard source iteration. 

iQMC treats the scattering and fission terms as internal fixed sources, and thus the Monte Carlo transport sweep is reduced to a particle ray trace which provides a well-suited application for Quasi-Monte Carlo. Quasi-Monte Carlo is the use of low-discrepancy sequences in place of pseudo-random number generators and provides a more efficient sampling of the phase space for a given number of particles. The improved sampling technique provides a theoretical $O(N^{-1})$ convergence compared to the  $O(N^{-1/2})$ of pseudo-random numbers. 

In each iQMC sweep, particles are emitted with an initial position and direction assigned from the LDS and a statistical weight calculated from a piecewise-constant source over a given mesh. $N$ particles are created and traced out of the volume, tallying the scalar flux with a path-length tally estimator. After a complete sweep, the new scalar flux approximation is sent to the iterative solver to update the source strength. iQMC was shown to achieve  $O(N^{-1})$ across multiple test problems, however was also observed to plateau in convergence in some problems \cite{pasmann2022quasi, pasmann2023iQMC}.

Previous iQMC results utilized a uniform Cartesian grid with a piecewise-constant source. Similar to ``teleportation error'' in Implicit Monte Carlo (IMC) methods \cite{Wollaeger2016}, the spatial discretization and piecewise-constant source can lead to a significant spatial error that limits convergence of the overall method. Taking concepts from IMC, we have developed a history-based discontinuous piecewise-linear source tilting scheme to reduce spatial error in iQMC. The source tilting method is described below and afterward we present results from a fixed-source 2D reactor-like problem adapted from the Takeda-1 Benchmark problem \cite{Takeda1991}.

%%%%%%%%%%%%%%%%%%%%%%%%%%%%%%%%%%%%%%%%%%%%%%%%%%%%%%%%%%%%%%%%%%%%%%%%%%%%%%%%
\section{METHODOLOGY}
\subsection{Piecewise-Constant Scheme}
In a 2-dimensional example, previous iQMC studies utilized a piecewise-constant flux approximation
\begin{equation}
    \phi_{\mathrm{Constant}} = a_{i,j},
\end{equation}
where $i$ and $j$ denote the 2-dimensional spatial indices and $a_{i,j}$ is the cell-averaged flux
\begin{equation}\label{eq:aij}
    a_{i,j} = \frac{1}{\Delta x_{i} \, \Delta y_{j}} \int_{y_{j-1}}^{y_{j}}\int_{x_{i-1}}^{x_{i}} \phi(x,y) \, dx \, dy .
\end{equation}
Given a path-length tally estimator with path length $S$, total cross section $\Sigma_t$, and continuous particle weight capture with initial weight $w_0$, the resultant tally is
\begin{equation}
    \int_{0}^{S} w_0 e^{-\Sigma_t S^\prime} \, dS^\prime = w_0 \left(\frac{1-e^{-\Sigma_t S}}{\Sigma_t} \right) .
\end{equation}
% $x\in[x_{i-1},x_i]$, and $y\in[y_{j-1},y_j]$. 
% \begin{equation}
%     \Delta x_j = x_j - x_{j-1}.
% \end{equation}
\subsection{Piecewise-Linear Scheme}
In our discontinuous linear scheme, the scalar flux is now represented as
\begin{multline} \label{eq:linear_scheme}
    \phi_{\mathrm{Linear}} = a_{i,j} + b_{i,j} (x-x_{\mathrm{mid},i}) 
    \\ 
    + c_{i,j} (y-y_{\mathrm{mid},j}) + d_{i,j} (x-x_{\mathrm{mid},i})(y-y_{\mathrm{mid},j}),
\end{multline}
where $x_{\mathrm{mid},i}$ and $y_{\mathrm{mid},j}$ represent the cell midpoint. The $a_{i,j}$ term is the same from the piecewise-constant scheme in Eq.~\ref{eq:aij}, while the $b_{i,j} \,$, $c_{i,j} \,$, $d_{i,j} \,$ terms respectively represent the linear flux-tilt in the $x$, $y$, and $xy$ directions, 
where the linear terms $b_{i,j} \,$ and $c_{i,j} \,$ are 
\begin{equation}
    b_{i,j} = \frac{12}{\Delta x_{i}^3 \Delta y_{j}} \int_{y_{j-1}}^{y_{j}}\int_{x_{i-1}}^{x_{i}} \left( x-x_{\mathrm{mid},i} \right)\phi(x,y) \, dx \, dy ,
\end{equation}
and
\begin{equation}
    c_{i,j} = \frac{12}{\Delta x_{i}\Delta y_{j}^3} \int_{y_{j-1}}^{y_{j}}\int_{x_{i-1}}^{x_{i}} \left( y-y_{\mathrm{mid},j} \right)\phi(x,y) \, dx \, dy , 
\end{equation}
and the bilinear term is 
\begin{multline}
    d_{i,j} = \frac{12}{(\Delta x_{i} \Delta y_{j})^3} \int_{y_{j-1}}^{y_{j}}\int_{x_{i-1}}^{x_{i}} \left( x-x_{\mathrm{mid},i} \right)
    \\
    \left( y-y_{\mathrm{mid},j} \right)\phi(x,y) \, dx \, dy , 
\end{multline}
The new terms can be similarly tallied as
\begin{equation}
     \int_{0}^{S} w_0 e^{-\Sigma_t S^\prime} \left[ \left(x_0 + \mu_x S^\prime\right)-x_\mathrm{mid} \right] \, dS^\prime, 
\end{equation}
for the $b_{i,j} \,$ and $c_{i,j} \,$ terms (swapping x-variables for y-variables respectively) and
\begin{multline}
     \int_{0}^{S} w_0 e^{-\Sigma_t S^\prime} \left[ \left(x_0 + \mu_x S^\prime\right)-x_\mathrm{mid} \right]
     \\
     \left[ \left(y_0 + \mu_y S^\prime\right)-y_\mathrm{mid} \right] \, dS^\prime, 
\end{multline}
for  $d_{i,j} \,$, where $x_0$ and $y_0$ denote particle initial position associated with the initial weight $w_0$.

%%%%%%%%%%%%%%%%%%%%%%%%%%%%%%%%%%%%%%%%%%%%%%%%%%%%%%%%%%%%%%%%%%%%%%%%%%%%%%%%
\section{TEST PROBLEM}
\subsection{2D Fixed-Source Reactor Problem}
\begin{figure}[p!] % replace 't' with 'b' to force it to be on the bottom
  \centering
  \includegraphics[width=0.48\textwidth]{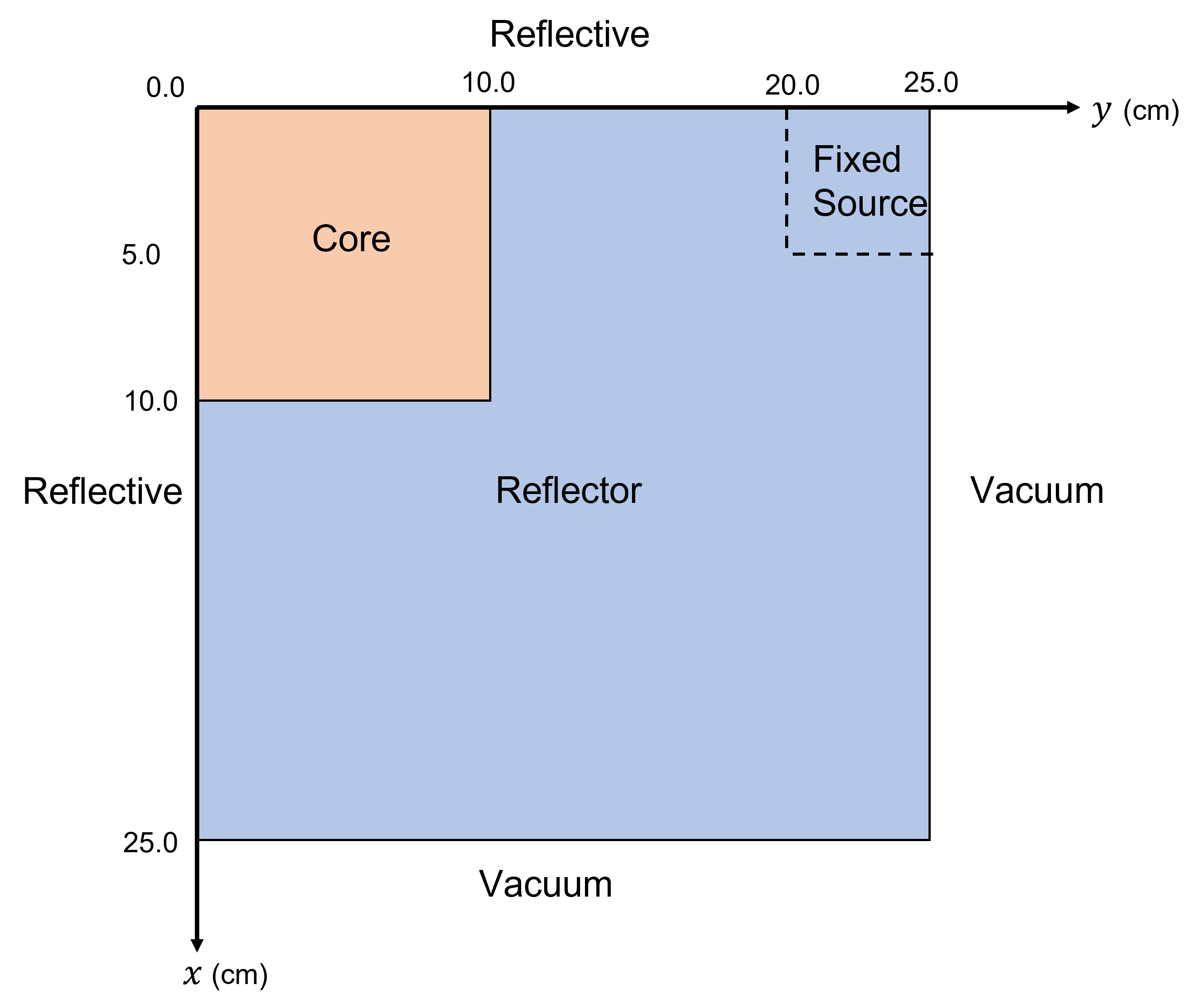}
  \caption{Core configuration of the test problem inspired by the Takeda-1 Benchmark~\cite{Takeda1991}.}
  \label{fig:setup}
\end{figure}

To evaluate the effect of using the proposed linear source, we've designed a 2-dimensional, 2-group, fixed-source reactor problem inspired by the Takeda-1 Benchmark problem \cite{Takeda1991} and using the same cross sections. The problem features a core region surrounded by reflective boundary conditions and reflector material, making the system near critical ($k=0.96883 \pm 0.00049$). A fixed source was placed outside the fuel in the reflector region.  Figure~\ref{fig:setup} depicts the problem setup. A reference scalar flux result is obtained from a high-fidelity Monte Carlo simulation generated using the Monte Carlo code MC/DC~\cite{variansyah_mc23_mcdc} with $10^{10}$ particle histories.

\begin{figure}[p!] % replace 't' with 'b' to force it to be on the bottom
  \centering
  \includegraphics[width=0.48\textwidth]{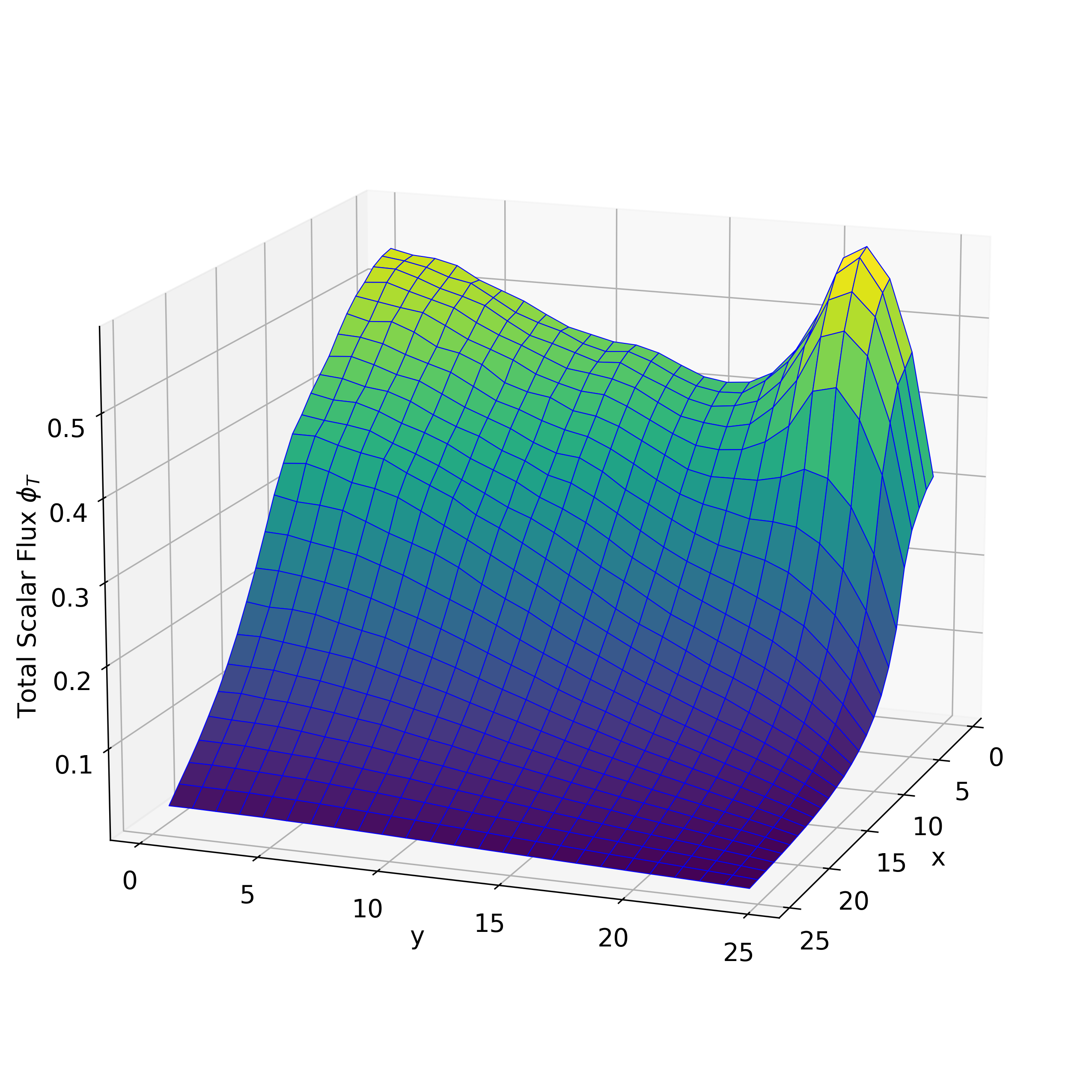}
  \caption{Total scalar flux results from iQMC simulation with source tilting and $N=2\times10^6$ particles per iteration.}
  \label{fig:flux}
\end{figure}

\begin{figure}[p!]
  \centering
  \includegraphics[width=0.48\textwidth]{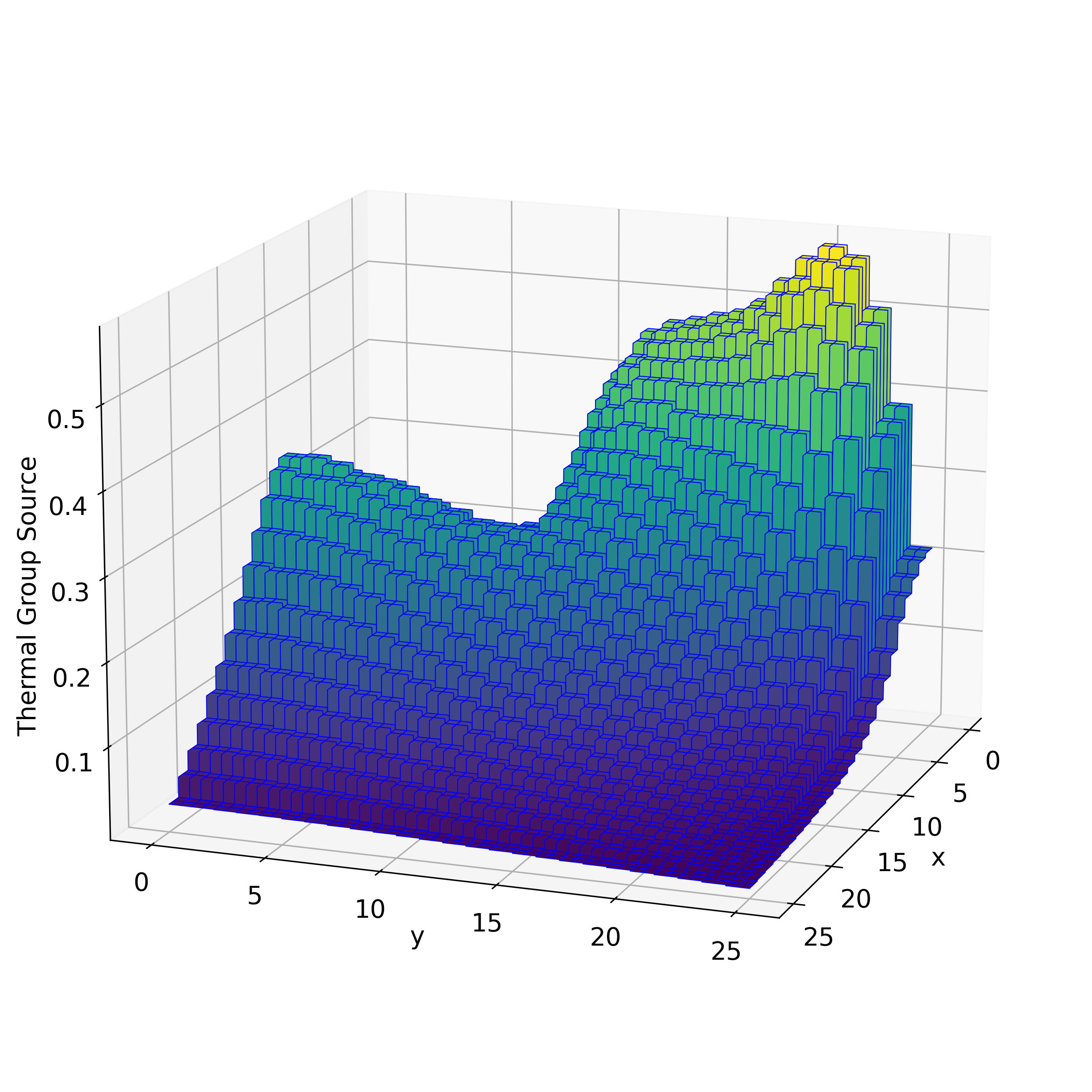}
  \caption{Final piecewise-constant thermal group source results from $N=2\times10^6$ particles per iteration.}
  \label{fig:constant_slow}
\end{figure}

\begin{figure}[p!]
  \centering
  \includegraphics[width=0.48\textwidth]{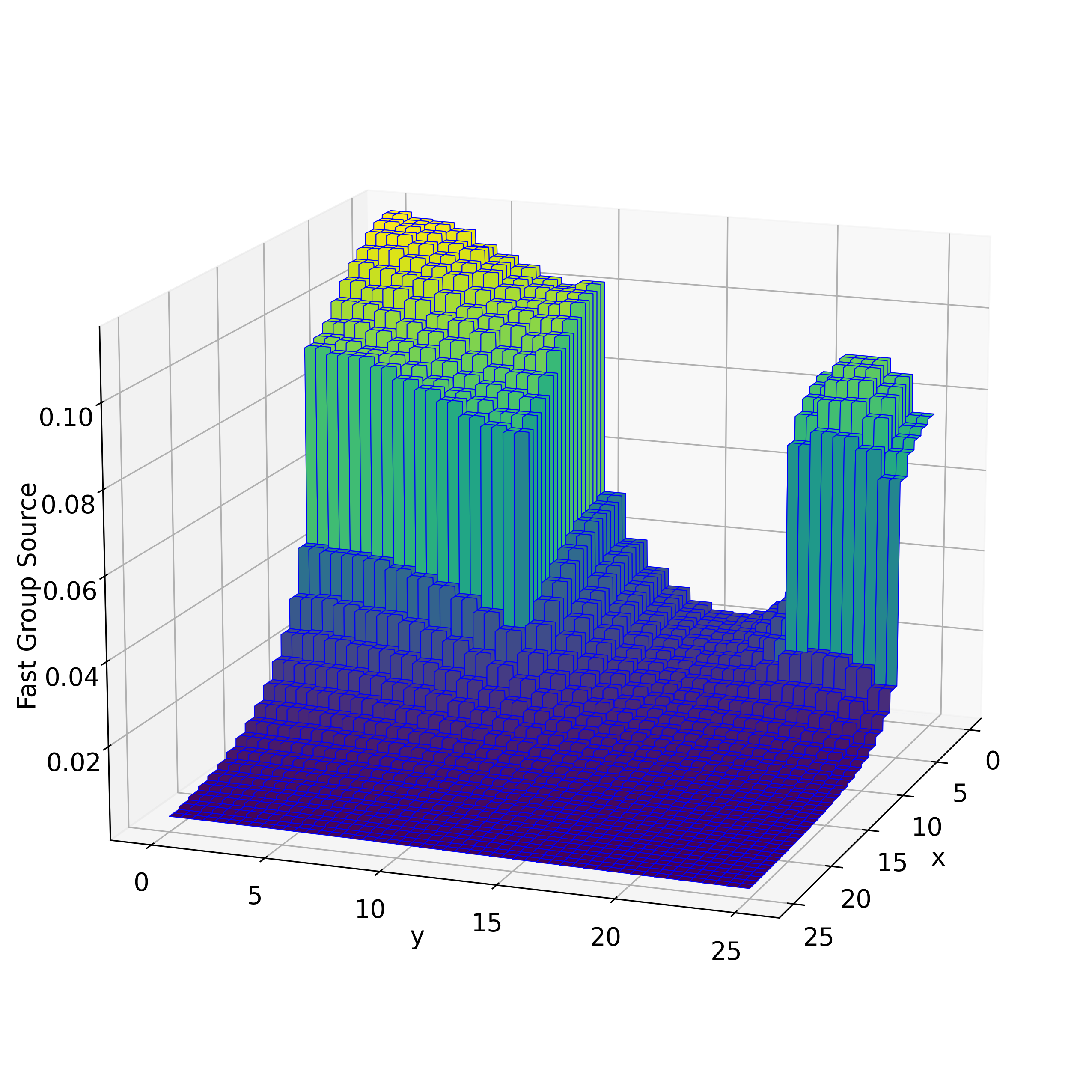}
  \caption{Final piecewise-constant fast group source results from $N=2\times10^6$ particles per iteration.}
  \label{fig:constant_fast}
\end{figure}

\begin{figure}[p!]
  \centering
  \includegraphics[width=0.48\textwidth]{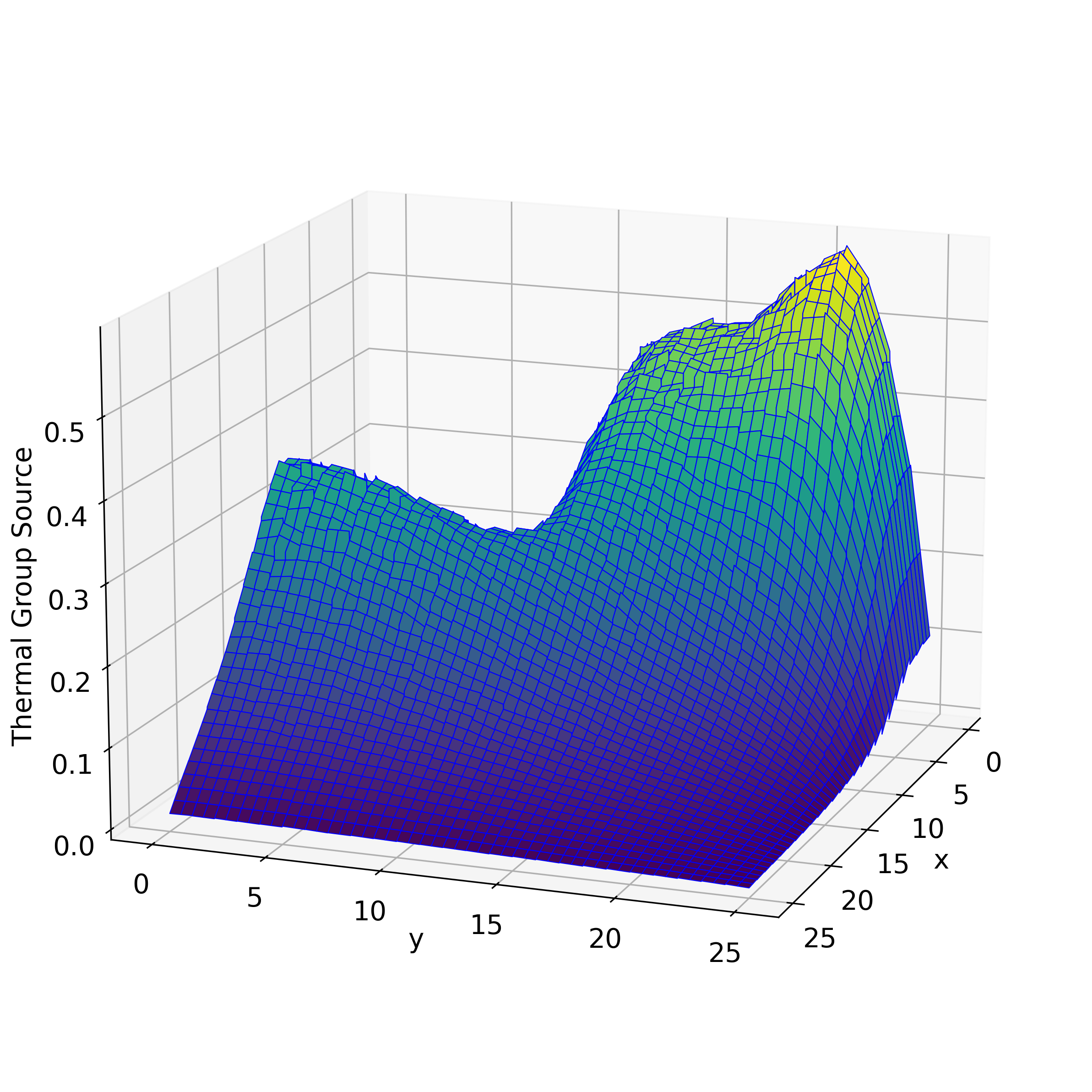}
  \caption{Final piecewise-linear thermal group source results from $N=2\times10^6$ particles per iteration.}
  \label{fig:tilted_slow}
\end{figure}

\begin{figure}[p!]
  \centering
  \includegraphics[width=0.48\textwidth]{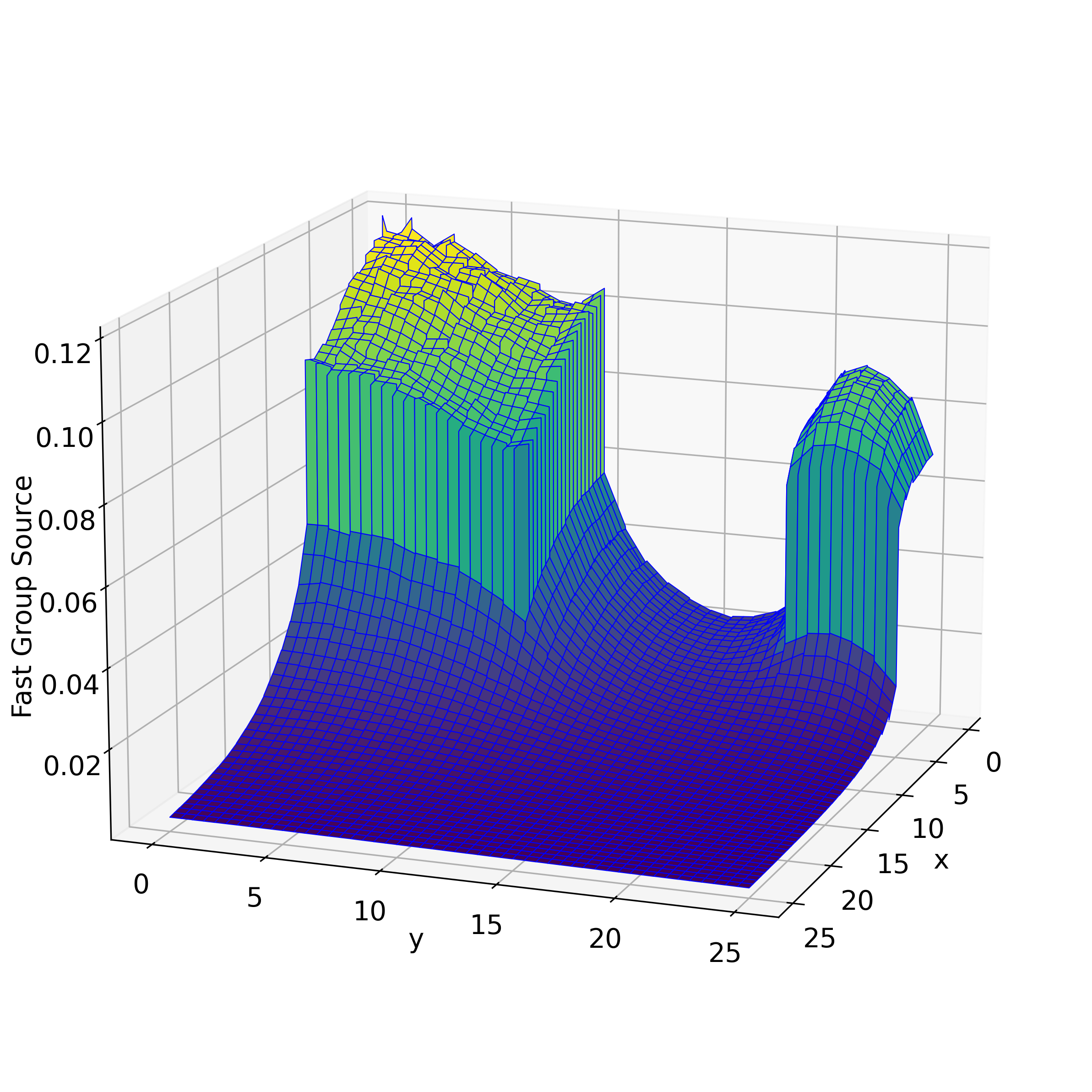}
  \caption{Final piecewise-linear fast group source results from $N=2\times10^6$ particles per iteration.}
  \label{fig:tilted_fast}
\end{figure}

\begin{figure}[ht!]
  \centering
  \includegraphics[width=0.48\textwidth]{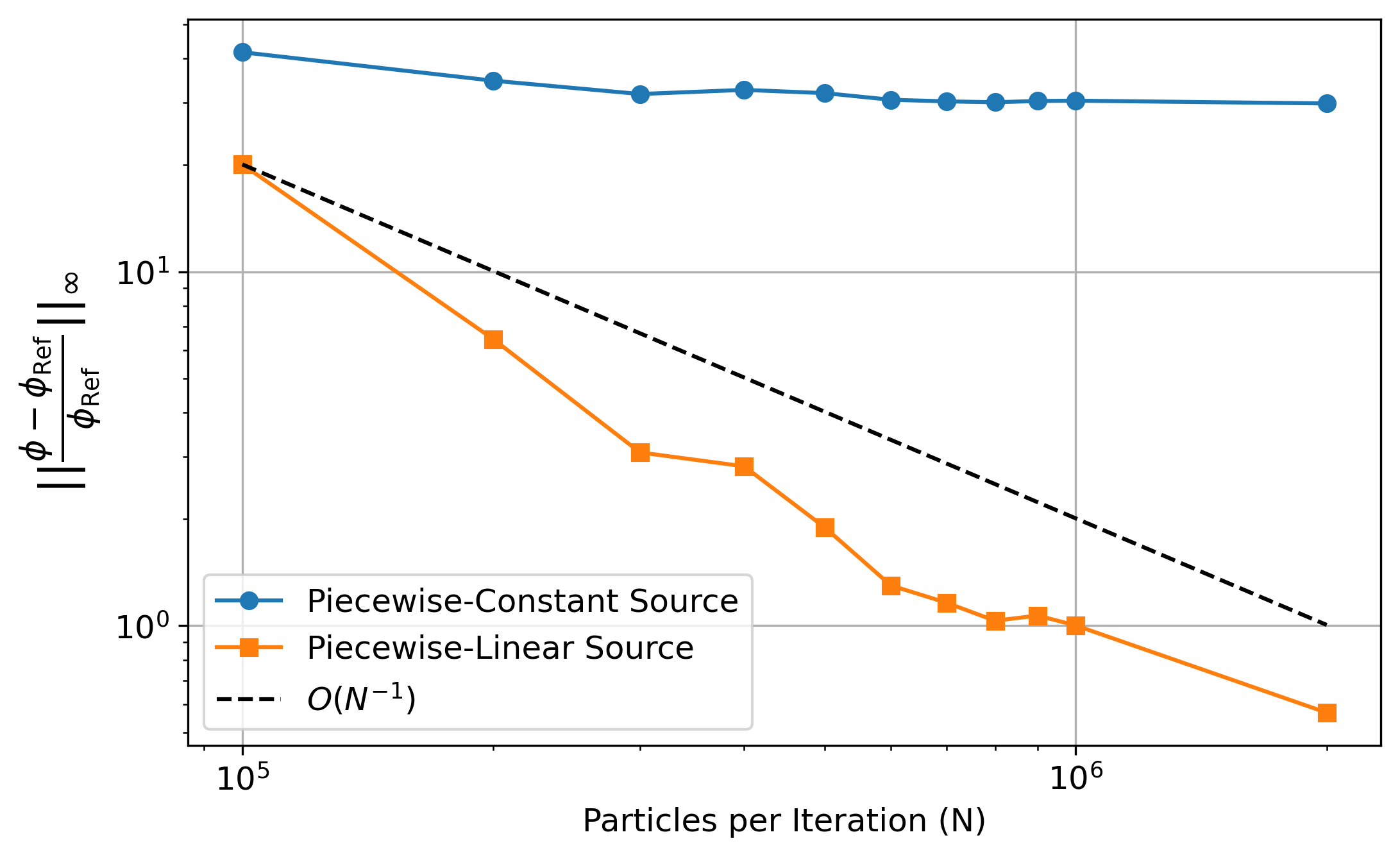}
  \caption{Scalar flux convergence as a function of particle histories per iteration ($N$). With a piecewise-constant source the error plateaus almost immediately. However, with a linear approximation iQMC is able to converge at the expected rate of $O(N^{-1})$.}
  \label{fig:convergence}
\end{figure}

The iQMC simulations were run with a $25\mathrm{X}25$ uniform mesh similar to other benchmark results of the Takeda-1 problem \cite{Takeda1991}. The Halton Sequence was used to generate particle positions and angles, while GMRES was used to iterate to $\Delta \phi/\phi \le 1\times10^{-9}$. Figure~\ref{fig:flux} shows the total (summed across groups) scalar flux from a resulting iQMC simulation with source tilting. Figures~\ref{fig:constant_slow} and \ref{fig:constant_fast} show the source strength from a piecewise-constant simulation, while Figures~\ref{fig:tilted_slow} and \ref{fig:tilted_fast} depict the source strength from a piecewise-linear simulation. Finally, Figure~\ref{fig:convergence} displays the $L_\infty$-norm of the scalar flux relative error for iQMC simulations with and without source tilting as a function of the number of particles per iteration.

\clearpage

%%%%%%%%%%%%%%%%%%%%%%%%%%%%%%%%%%%%%%%%%%%%%%%%%%%%%%%%%%%%%%%%%%%%%%%%%%%%%%%%

\section{CONCLUSIONS}
We have developed a history-based linear discontinuous source tilting scheme for the iterative Quasi-Monte Carlo (iQMC) method. The source tilting technique was shown to reduce spatial error in a 2-dimensional, 2-group, reactor problem. Figure~\ref{fig:convergence} shows that the relative error plateaus from piecewise-constant simulations, while the piecewise-linear results are able to converge at the expected $O(N^{-1})$. It is unlikely however, that piecewise-linear source tilting completely eliminates the spatial discretization error. More tests are needed to further explore the limitations of the method and to evaluate performance on more difficult 3D problems.  

%%%%%%%%%%%%%%%%%%%%%%%%%%%%%%%%%%%%%%%%%%%%%%%%%%%%%%%%%%%%%%%%%%%%%%%%%%%%%%%%
% \appendix
% \section{Appendix}

% Numbering in the appendix is different:
% \begin{equation} \label{eq:appendix}
%   2 + 2 = 5\,.
% \end{equation}
% and another equation:
% \begin{equation} \label{eq:appendix2}
%   a + b = c\,.
% \end{equation}

%%%%%%%%%%%%%%%%%%%%%%%%%%%%%%%%%%%%%%%%%%%%%%%%%%%%%%%%%%%%%%%%%%%%%%%%%%%%%%%%

\section{Acknowledgments}
This work was funded by the Center for Exascale Monte-
Carlo Neutron Transport (CEMeNT) a PSAAP-III project
funded by the Department of Energy, grant number: DE-
NA003967 and the National Science Foundation, grant number DMS-1906446.

%%%%%%%%%%%%%%%%%%%%%%%%%%%%%%%%%%%%%%%%%%%%%%%%%%%%%%%%%%%%%%%%%%%%%%%%%%%%%%%%

\bibliographystyle{ans}
\bibliography{ans23}
\end{document}